%% file: stm13_AF.tex
\documentclass[b5paper,twoside]{stm13b5}

\usepackage[pdftex]{graphicx} 

\usepackage[utf8]{inputenc}
\usepackage[T1]{fontenc}
\usepackage{euscript}
\usepackage{amsfonts,amssymb,amsmath}
\usepackage{amsthm}
\usepackage{feynmp}
\usepackage{enumerate}
\usepackage{amscd}

\usepackage{hyperref}

\def\url#1{\texttt{#1}}

\begin{document}

\setcounter{equation}{0}
\setcounter{figure}{0}
\setcounter{section}{0}
\input{Author_13.tex}

\end{document}

%% file: Author_13.tex
% !TeX root=sample_STM_2013_b5.tex

\thispagestyle{plain}
\addcontentsline{toc}{subsection}{\numberline{}\hspace*{-15mm}A. Bob\'ak, M. Borovsk\'y, T. Lu\v{c}ivjansk\'y, M. \v{Z}ukovi\v{c}:
\emph{Tricritical Properties of Antiferromagnetic Ising Model on the Square Lattice}}

\markboth{\sc A. Bob\'ak {\it et al.}} {\sc Tricritical Properties of Antiferromagnetic Ising Model}

\STM

\title{Tricritical Properties of Antiferromagnetic Ising Model on the Square Lattice}

\authors{A. Bobák$^1$, M. Borovský$^1$, T. Lu\v{c}ivjansk\'y$^{1,2}$, M. Žukovič$^1$}

\address{$^{1}$
\v{S}af\'arik University,
Institute of Physical Sciences,\\
Park Angelinum 9, 040 01 Košice,
Slovak Republic\\
$^{2}$
Institute of Experimental Physics, Slovak Academy of Sciences,\\
Watsonova 47, 040 01 Ko\v{s}ice, Slovak Republic}

\bigskip

\begin{abstract}

\hspace*{0.5cm}The Ising square lattice model with nearest-neighbor (nn) interactions $(J_1)$ is one of the few exactly solvable models
 \cite{Bax07}.
 Adding next-neareast-neighbor (nnn) interactions $(J_2)$ or a magnetic field (or both) leads to
the non solvability of the model and only some approximate solutions are possible. In this brief report
we will review some results obtained within effective field theory. We will show that  
besides second-order transitions there are also lines of first-order transitions and the coordinates of
 tricritical points are calculated.
 \end{abstract}

\vspace*{3mm}
 
\begin{figure}
 \centerline{\includegraphics[width=0.4\textwidth]{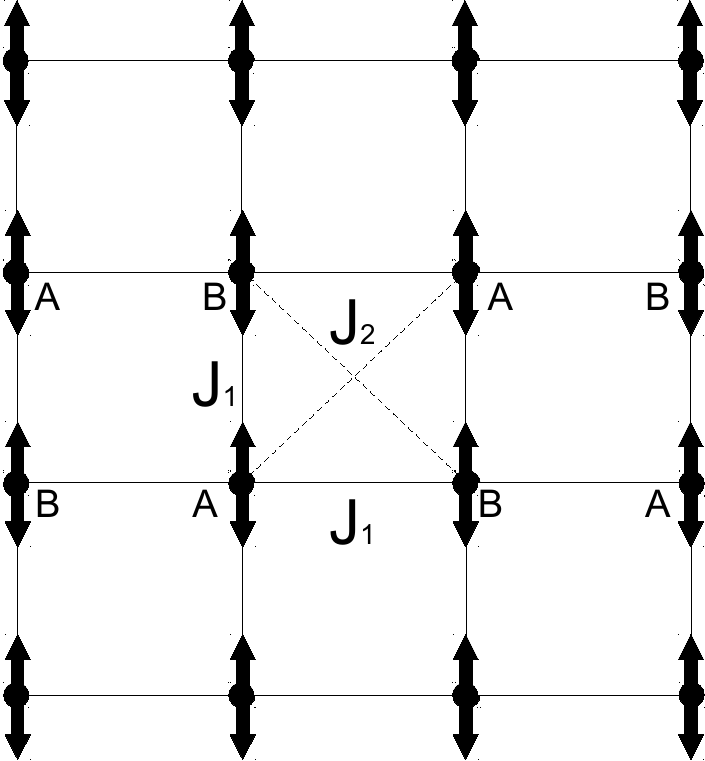}} % r_higgs_fig1.eps
 \caption{Geometrical representation of the model Hamiltonian (\ref{ham}). The whole lattice can be divided into two sublattices
    denoted here as A and B. Interactions between the neighboring spins within the given sublattice is $J_2$ and interactions between
    neighboring spins from different sublattices is $J_1$.}
 \label{fig:struct}
 \end{figure}

The Hamiltonian of the Ising model with competing antiferromagnetic (AF) interactions between nn $(J_1 < 0)$ and nnn $(J_2 < 0)$ spins in zero
external field is 
\begin{equation}
\label{ham}
H = -J_1\sum_{nn} \sigma_i \sigma_j - J_2 \sum_{nnn} \sigma_i \sigma_j,
\end{equation}
where classical spin variable $\sigma_i$ can acquire two values $\pm 1$, and
summations runs over corresponding pairs of interacting spins. For better visualization
 the geometrical structure of the model is depicted on the Fig. \ref{fig:struct}. 
 From an experimental point of view, such model could describe
quasi-two-dimensional anisotropic AF like K$_2$CoF$_4$ compound at least within a fair 
approximation. Since the existence of nnn interactions and the magnetic field may give rise to other phases with different types of phase 
transitions and multicritical points, we restrict ourselves to the AF interactions with zero magnetic field.
In such model it is also expected presence of so-called superantiferromagnetic (SAF) - striped phase.

Our main aim is to determine the phase diagram in the parameter space $(\alpha,t)$ (defined below, see \ref{eq:def_alpha})
using an effective field theory based on succesive use of one-, two-, and four-spin clusters. Used approach is based on the differential
 operator technique \cite{HoKa78} introduced into exact Ising spin identities  and has already been successfully applied to a variety of
 spin-$\frac {1}{2}$ and higher spin problems.
 Of course, the effective field calculation cannot give any precise information about the value of critical exponents.
The starting point of the effective field theory is the so-called set of correlation identities.
By standard procedure involving use of van der Waerden identity and truncation procedure it is
possible to get closed set of equations for spin-correlations functions\cite{TF82}. We will restrict ourselves to the simplest possible
truncation scheme
\begin{equation}
   \langle \sigma_i \cdot \sigma_j \cdots \rangle \approx \langle \sigma_i\rangle \cdot
   \langle \sigma_j\rangle \cdots \quad (i\neq j\neq \cdots).
\end{equation}
In order to take into account effects of frustration within the present effective-field theory, it is 
necessary to consider at least two-spin cluster. By relatively straightforward calculation 
the normalized staggered magnetization
 $m_s$ associated with the antiferromagnetic two-spin cluster is given by  
 \begin{eqnarray}
\label{magaf}
m_s &=& \Big[A_x(1) A_y(2) + B_x(1) B_y(2) + m_B \Big(A_x(1)B_y(2)+A_y(2)B_x(1)\Big)\Big]^2 \nonumber \\
& & \times \Big[A_y(1) A_x(2) + B_y(1) B_x(2) + m_A \Big(A_y(1)B_x(2)+A_x(2)B_y(1)\Big)\Big]^2 \nonumber \\
& & \times \Big(A_x(1) + m_B B_x(1)\Big)\Big(A_y(1) + m_A B_y(1)\Big) \nonumber \\
& & \times \Big[\Big(A_x(2) + m_A B_x(2)\Big)\Big(A_y(2) + m_B B_y(2)\Big)\Big]^2f_s (x,y)\Big|_{x=0,y=0}, 
\end{eqnarray}
where $m_s=(m_A-m_B)/2$ represents an order parameter for AF phase,
  $A_\mu(\nu) = \cosh(J_\nu D_\mu)$, $B_\mu(\nu) = \sinh(J_\nu D_\mu)$ ($\nu = 1, 2$), $D_\mu \equiv \partial/\partial \mu$.
The function $f_s(x, y)$ is defined as follows
\begin{equation}
\label{func}
f_s (x, y) = \frac{\sinh\beta(x-y)}{\cosh\beta(x-y) + e^{2\beta J_1}\cosh\beta(x+y)},
\end{equation} 
where $\beta = 1/{\rm k_B}T$ is reduced inverse temperature.
 Equation of state (\ref{magaf}) is quite superior to that obtained from the standard mean-field theory, since in the
 present framework relations like $(s_g^\alpha)^{2r} = 1$ and $(s_g^\alpha)^{2r+1} = s_g^\alpha$, for all $r$, are taken exactly into account through 
the van der Waerden identity, $\exp(as_g^\alpha) = \cosh(a) + s_g^\alpha \sinh (a)$. On the other hand in the usual molecular-field theory not only
multi-spin correlations are neglected but so are self-correlations of spin variables. 

It is convenient to
introduce dimensionless effective parameters $t$ and $\alpha$ by the following relations
  \begin{equation}
      \alpha = \frac{J_2}{|J_1|},\quad
      t = \frac{1}{\beta |J_1|}.
      \label{eq:def_alpha}
  \end{equation}
 Since for the antiferromagnetic $J_1-J_2$ model at zero magnetic field the following symmetry $m_s \longleftrightarrow -m_s$
for the staggered magnetization holds, Eq.~(\ref{magaf}) can be recast in the form 
\begin{equation}
\label{magm_s}
m_s = \sum_{{n}=0}^{4} K_{2n+1}^{AF} m_s^{2n+1}, 
\end{equation}
where only the odd coefficients $K_{2n+1}^{AF}$ are present. These coefficients depend on $t$ and $\alpha$ and can be easily calculated 
using symbolic calculations by direct employ of translational formula
\begin{equation}
  \exp (\lambda D_x + \gamma D_y) f_s(x, y) = f_s(x+\lambda, y+\gamma).
\end{equation}
 Because the final expressions for 
 coefficients $K_{2n+1}^{AF}$ are quite lengthy, their explicit form is omitted. We note also that in obtaining Eq.~(\ref{magm_s}) we have made use of the 
fact that $f_s(x, y) = - f_s(-x, -y)$ and therefore only odd differential operator functions remain.\\   
\hspace*{0.5cm}In order to determine the phase diagram of the antiferromagnetic $J_1-J_2$ model, we should solve Eq.~(\ref{magm_s}) 
for a given value of the frustration parameter $\alpha$ and look for the temperature at which the magnetization (order parameter) $m_s$ 
goes to zero. However, for a some value of $\alpha$, the order parameter goes to zero discontinously, accordingly the transition becomes first 
order. To analyze such transition, one needs to calculate the free energy for the antiferromagnetic and paramagnetic phases
 and to find a point of intersection. Because the expression for the free energy in this effective-field theory does not exist, it 
will be extrapolated with the help of the relation for the equilibrium value of the order parameter (\ref{magm_s}) as follows (\cite{SB82}):
\begin{equation}
\label{freeenergy}
F^{AF}(t, \alpha, m) = F_0(t, \alpha) + \frac{1}{2}\Big(1-\sum_{{n}=0}^{4} \frac{K_{2n+1}^{AF}}{n+1}{m^{2n}}\Big)m^2, 
\end{equation}
where $F_0(t,\alpha)$ is a free energy of the disordered (paramagnetic) phase and $m$ is the order parameter which takes the value $m_s$ at thermodynamic
 equilibrium. We note that relation (\ref{freeenergy}) corresponds to a Landau free energy expansion in the order parameter truncated in the $m^{10}$
 term. \\  
\hspace*{0.5cm}Thus the equilibrium magnetization is the value of the order parameter, which minimizes the free energy given by the Eq.~(\ref{freeenergy}).
 Using the equilibrium condition  
\begin{equation}
\label{eqcondition}
\frac{\partial F^{AF}(T, \alpha, m)}{\partial m}\Big|_{m=m_s} = 0, 
\end{equation}
we recover the Eq.~(\ref{magm_s}) for the equilibrium magnetization. Then a critical temperature and a tricritical point, at which the phase transition 
changes from second order to first order, are determined as follows: (i) second-order transition line when $1-K_1^{AF} = 0$ and $K_3^{AF} < 0,$ and (ii) 
tricritical point when $1-K_1^{AF} = 0$, $K_3^{AF} = 0,$ if $K_5^{AF} < 0$. It is worth noticing that if we use this relations to obtain the critical and 
tricritical points, the results coincide with those obtained from Eq.~(\ref{magm_s}). We believe that this justifies our procedure. However, the 
first-order phase transition line is evaluated by solving simultaneously two transcendental equations, namely (\ref{eqcondition}) and
 $F^{AF}(t,\alpha , m) = F_0(t,\alpha)$, which corresponds to the point of intersection of the free energy antiferromagnetic and paramagnetic phases. \\     
\hspace*{0.5cm} Analogous procedure could also be performed for the superantiferromagnetic phase. After some algebraic
manipulations it can be shown that the corresponding equation of state is 
\begin{eqnarray}
\label{magsaf}
m_s&=& \Big[A_x(1) A_y(2) + B_x(1) B_y(2) + m_A \Big(A_x(1)B_y(2)+A_y(2)B_x(1)\Big)\Big]^2 \nonumber \\
& & \times \Big[A_y(1) A_x(2) + B_y(1) B_x(2) + m_B \Big(A_y(1)B_x(2)+A_x(2)B_y(1)\Big)\Big]^2 \nonumber \\
& & \times \Big(A_x(1) + m_B B_x(1)\Big)\Big(A_y(1) + m_A B_y(1)\Big) \nonumber \\
& & \times \Big[\Big(A_x(2) + m_B B_x(2)\Big)\Big(A_y(2) + m_A B_y(2)\Big)\Big]^2f_s (x,y)\Big|_{x=0,y=0}. 
\end{eqnarray}
Thus, using of the same sort of approximate scheme as before, we obtain numerically the second- and first-order transition lines including
 the tricritical point between the superantiferromagnetic and paramagnetic phases. \\
\hspace*{0.5cm} One can of course imagine a continuation of the this process with considering larger and larger clusters.
As a consequence better results for thermodynamic properties are expected. Because of natural limitations of computer power one
cannot expect to solve this problem exactly (which would basically correspond to the infinite cluster) within the used method. 
Therefore, the next step is a cluster with four spins containing the information of the lattice topology.
Using the same procedure as before for the two-spin cluster, one determines the second- and first-order transition line including tricritical 
point between the ordered (antiferromagnetic or superantiferromagnetic) and paramagnetic phases. 

\begin{figure}
  \centering
   \includegraphics[width=0.5\textwidth]{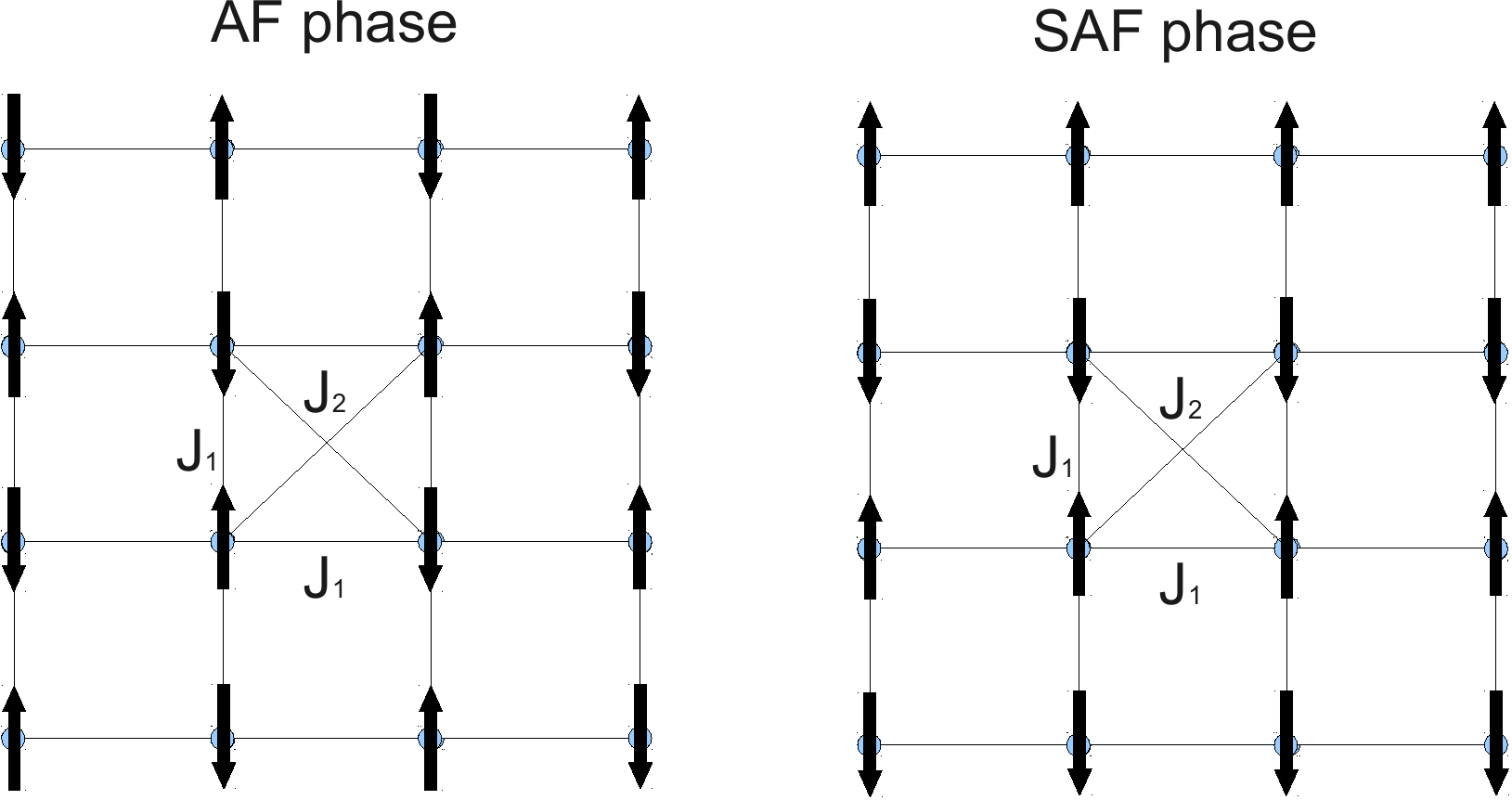}
   \caption{Ground state configuration for the model }
   \label{Fig:OS}  
\end{figure}
   
\begin{figure}
 \centerline{\includegraphics[width=0.75\textwidth]{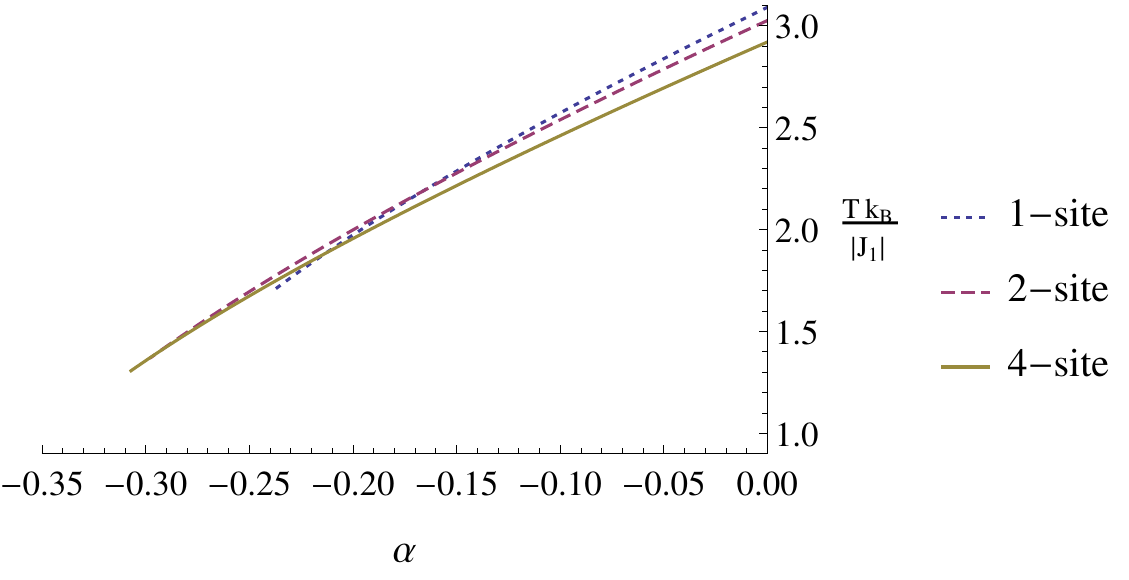}} % r_higgs_fig1.eps
 \caption{Second order transition line between antiferromagnetic and paramagnetic phase.}
 \label{fig:hmass_vs_tan}
 \end{figure}

Let's discuss basic features of the phase diagram.
To get some insight into the problem we start with analysis of the ground state ($t=0$).
The expected ordered states are graphically represented on the Fig. \ref{Fig:OS} and their energies are
\begin{equation}
      \frac{E_{SAF}}{N|J_1|} = 2\alpha, \quad \frac{E_{AF}}{N|J_1|} = -2(1+\alpha)  .
\end{equation}
By direct comparison we see, that for $\alpha=-0.5$ there should be a phase transition from
the AF phase to the SAF phase. 
Next by numerical analysis of equations (\ref{magaf}) and (\ref{magsaf}) we were able to obtain
second order transition lines between AF and para phase and also
between SAF and para phase. Because of lack of space we will present here only results for the former case, which
is depicted on the Fig.\ref{fig:hmass_vs_tan}. 
As can be readily seen all used approximations lead to the prediction of existence of tricritical point. 
We obtained the similar picture also for the superantiferromagnetic-paramagnetic transition. 
%\fontsize{10pt} 
\begin{table}
\centering
\renewcommand{\arraystretch}{2}
       \begin{tabular}{ | c |c | c | c |}
  \hline
     & 1-site & 2-site & 4-site \\ \hline 
  $t_{tcp}^{AF}$ & $1.7002$ & $1.3720$ & $1.3066$  \\
  $\alpha_{TCP}^{AF}$ & $-0.2383$ & $-0.2973$ & $-0.3070$ \\ \hline
$t_{tcp}^{SAF}$ & $3.2207$ & $2.2594$ & $2.4619$  \\
  $\alpha_{tcp}^{SAF}$ & $-1.3276$ & $-0.9991$ & $-1.0387$ \\ \hline
  $t_{cp}$ & $3.0898$ & $3.0250$ & $2.9197$ \\ \hline  
 \end{tabular}
  \caption{Coordinates of tricritical points and critical points for $\alpha=0$ in various approximations.}
  \label{tab:results}  
\end{table}
To have more quantitative comparison we have listed coordinates for the tricritical points in the
Tab. \ref{tab:results}. Note also critical temperatures $t_c$ for the case $\alpha=0$. The relative decreasing
of the critical temperature is in agreement with the exact value $t_c^{exact}=2.2692$ for the given lattice \cite{Domb74}.

Effective field theory doesn't capture fluctuations in order parameter. It is well known that fluctuations are greatly pronounced in the
critical region. Actually in some cases such fluctuations could lead to change 1-order phase transition into 2-order 
phase transition as is
e.g. the case for random Ising model \cite{AW90}. Such possibility could not be excluded by the effective field theory. Therefore
 more sophisticated methods are needed, such as Monte Carlo simulations \cite{KHM11,KaHo12}. This method however have its
own deficiencies such as sign problem in the region $\alpha \approx -0.5$. 
  Our next steps will be in automatization of algorithms to deal with $N^2$-site cluster approximation ($N=3,4,\ldots$) and in comparison
of obtained results with Monte Carlo simulations.

% \begin{figure}
% \centerline{\includegraphics[width=0.65\textwidth]{r_higgs_fig20.pdf}} % r_higgs_fig2.eps
% \caption{Regions of the parameter space excluded by the non-observation of the Higgs boson for different values of $\varepsilon$}
% \label{fig:param_space}
% \end{figure}

\section*{Acknowledgements}
This work was supported by the Scientific Grant Agency of Ministry of Education of Slovak Republic (Grant VEGA No.1/0222/13 and
Grant No. 1/0234/12). 
This article was also created by implementation of the Cooperative phenomena and phase transitions in
nanosystems with perspective utilization in nano- and biotechnology
project No 26110230097. Funding for the operational research and development program was provided by the European Regional 
Development Fund. One of us (MB) was also
supported by the Faculty of Science UPJŠ (Grant ID. VVGS-PF-2013-94).